# Energy and Delay aware Physical Collision Avoidance in Unmanned Aerial Vehicles

Sihem Ouahouah [*||], Jonathan Prados-Garzon [**], Tarik Taleb[||] and Chafika Benzaid[§]
[||] Aalto University, Espoo, Finland. Emails: tarik.taleb@aalto.fi, sihem.ouahouah@aalto.fi
[*] Ecole nationale Suprieure d'Informatique, Algiers, Algeria. Email: s_ouahouah@esi.dz
[**] Department of Signal Theory, Telematics, and Communications, Univ. of Granada, Granada, Spain. Email: jpg@ugr.es
[§] LSI, USTHB, Algiers, Algeria. Email: cbenzaid@usthb.dz

***Abstract*—Several solutions have been proposed in the literature to address the Unmanned Aerial Vehicles (UAVs) collision avoidance problem. Most of these solutions consider that the ground controller system (GCS) determines the path of a UAV before starting a particular mission at hand. Furthermore, these solutions expect the occurrence of collisions based only on the GPS localization of UAVs as well as via object-detecting sensors placed on board UAVs. The sensors' sensitivity to environmental disturbances and the UAVs' influence on their accuracy impact negatively the efficiency of these solutions. In this vein, this paper proposes a new energy- and delay-aware physical collision avoidance solution for UAVs. The solution is dubbed EDC-UAV. The primary goal of EDC-UAV is to build in-flight safe UAVs trajectories while minimizing the energy consumption and response time. We assume that each UAV is equipped with a global positioning system (GPS) sensor to identify its position. Moreover, we take into account the margin error of the GPS to provide the position of a given UAV. The location of each UAV is gathered by a cluster head, which is the UAV that has either the highest autonomy or the greatest computational capacity. The cluster head runs the EDC-UAV algorithm to control the rest of the UAVs, thus guaranteeing a collision-free mission and minimizing the energy consumption to achieve different purposes. The proper operation of our solution is validated through simulations. The obtained results demonstrate the efficiency of EDC-UAV in achieving its design goals.**

## I. Introduction

Unmanned Aerial Vehicles (UAVs), popularly known as drones, have proven to be useful for a wide spectrum of applications over the past few decades [1]–[3]. Moreover, in recent years, new UAV applications in the civilian and commercial domains have emerged due to the continuous reduction in the price of UAVs and achievements in device miniaturization [4]. The range of UAV applications includes, among many others, traffic monitoring, load transport, environmental monitoring, disaster prevention, and capacity and coverage enhancement in wireless communications. Regarding the latter, UAVs are seen as part of the upcoming 5G mobile system. Their connectivity as well as the remote management of their traffic control based on cellular systems bring many interesting challenges that the research community is working on resolving [5]–[7].

In order to enable a number of computation-intensive services on board the UAVs (e.g., UAV-based video surveillance), and considering their limited computation resources, several solutions for offloading computation from UAVs have been also explored [8]. For applications where autonomous UAVs are required, each UAV is given general tasks, and it can accomplish them without human intervention. In such scenarios, one of the key prerequisites for UAVs is the sense-and-avoid capability equivalent to that of the human pilot [1].

A recent study has shown that the production of UAVs for personal and commercial use is proliferating, with global market revenue expected to increase by 34% and worth more than $2, 6 billion in this year and more than $11.2 billion by 2020 [9]. Almost two million UAVs will be produced in 2018, 39% more than the previous year [9]. With the expected growing number of UAVs, preventing collisions among UAVs could be a real challenge soon.

Several research works have tackled the collision avoidance problem in the context of ground vehicles [10], [11] as well as in the context of UAVs [12]–[23]. Most existing solutions assume the exact accuracy of the data captured by sensors and GPS. However, real experience shows that their precision could be easily affected by external conditions (e.g., weather conditions). Some solutions use cameras as sensors to detect the obstacles around UAVs [12], [20]. Nevertheless, the information provided by video cameras requires intensive processing to be translated into useful information to control UAVs [24], making cameras unsuitable for specific application scenarios. In the same way, the use of radar, as in [14], is not appropriate for small-scale UAVs due to its weight and size [24].

In this vein, this work proposes a novel solution for flying a swarm of UAVs (i.e., multiple movers problem [25]) within a bounded area (two dimensional horizontal plane -$2D_H$ -) towards a target destination, while ensuring collision avoidance. Our solution follows a global approach for collision avoidance. In the universal method, all possible collisions are evaluated and resolved at once [26]. Even though general methods exhibit high computational complexity, they may offer more robust solutions than the pairwise approaches, where the potential collisions are addressed sequentially in pairs [26]. Our solution minimizes the distance traveled by each UAV to reach its target destination (while ensuring no collision), thus helping to optimize the energy consumption and the overall system response time.Our solution considers each UAV has a GPS sensor on board to identify its position. To guarantee a collision-free operation, our solution defines a safety region for each UAV that accounts for the margin of

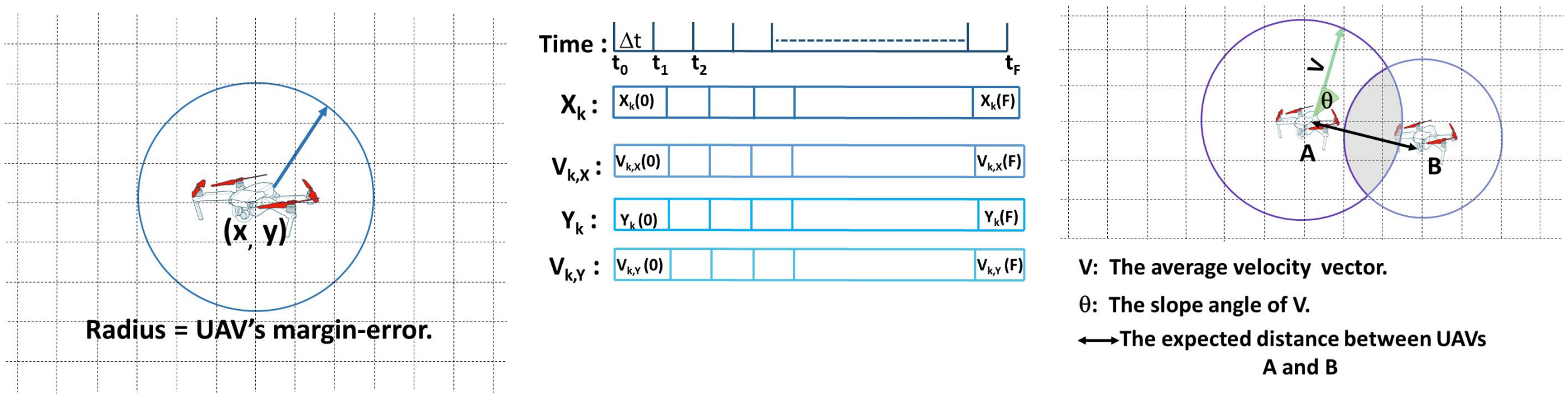

(a) UAV's margin-error. (b) UAV's slot time and parameters definition. (c) The expected collision zone between two UAVs.

Fig. 1. System model's description.

error in GPS localization [27]. Also, to enhance the UAVs' autonomy, we adopted a cluster architecture where the computation is offloaded to the cluster head (CH), elected based on the metrics defined in [28]. We assess the performance and validate the proper operation of our solution by means of simulations.

The remainder of this paper is organized as follows. Section II includes the system model and the formulation of the UAVs collision avoidance problem. Section III provides an overview of our solution and addresses the resolution of the optimization problem of the UAVs collision avoidance. Section IV presents some results to assess the performance and validate the correct operation of our solution. Finally, Section V draws the main conclusions.

## II. System Model and problem formulation

Let us assume a set of $K$ UAVs, where each UAV in the set is equipped with a GPS sensor. The GPS sensor exhibits a margin-error $r$, which is assumed to be known (see Fig. 1(a). We also assume that the UAVs fly within a two-dimensional area $X \times Y$. Initially, each UAV $k \in [1, K]$, has a fixed start-location $(S_X(k), S_Y(k))$, and a final target location $(E_X(k), E_Y(k))$. For the sake of simplicity, we assume that the trajectory of each UAV is a straight line, though the proposed solution can be easily adapted to more complex trajectories. We consider that the mission time $T$ is divided into a set of time slots, whose duration is denoted as $\Delta t$ as depicted in Fig. 1(b). To model the GPS margin error, we suppose that the UAV might be located within a circle of radius $r_k$ centered at the position of UAV $k$ reported by GPS module at time slot $t$. We also assume that the actual UAVs position within the circle obeys a uniform distribution. We consider that the occurrence of overlapping between two circles of two UAVs means that there is a probability of collision between them (see Fig. 1(c)).

For each $\Delta t$, our solution runs an algorithm to compute the next-position where the UAVs have to move without risk of collisions. The algorithm ensures that any couple of circles representing the potential locations of two UAVs are never overlapped, while minimizing the total distance traveled by each UAV. To that end, the angle between the UAVs' velocity vector and the target destination is minimized. The set of actions generated by the collision avoidance algorithm are interpreted as a set of commands (e.g., change trajectory, stop) that are subsequently sent to the UAVs to program their movements. Each action is composed of three parameters, $V_k$, $\theta_k$ and $\alpha_k$, which refer to the velocity, the angle slope, and the expected deviation angle of $V_k$, respectively $T$ is subdivided into a set of time slots, such that $t_0$ is the first time slot and $t_F$ is the last time slot. We define the variables $(X_u(t), Y_l(t))$ that specify the location of a UAV $u \in [1, K]$ at time slot $t \in \{t_1, t_2, \cdots t_F\}$. Formally, the process of seeking the optimal angle of deviation of UAVs velocity vector is modeled as follows:

$$min \sum_{k=1}^{K} |\alpha_k(t) - \theta_k(t)| \qquad (1)$$

s.t.

$$0 \leq |\alpha_k(t) - \theta_k(t)| \leq 2\pi \qquad (2)$$

$$X_k(t+1) = X_k(t) + V_k(t).\cos(|\alpha_k(t) - \theta_k(t)|) \times \Delta t \qquad (3)$$

$$Y_k(t+1) = Y_k(t) + V_k(t).\sin(|\alpha_k(t) - \theta_k(t)|) \times \Delta t \qquad (4)$$

$$\begin{aligned} B_k(\boldsymbol{x_c}, r_k) &= \{\boldsymbol{x}, \quad ||\boldsymbol{x} - \boldsymbol{x_c}||_2 \leq r_k\} \\ and \quad \boldsymbol{x_c} &= (X_k(t+1), Y_k(t+1)); \quad k \in [1, K] \end{aligned} \qquad (5)$$

$$\begin{aligned} dist(B_k, B_l) &= inf\{||\boldsymbol{x_k} - \boldsymbol{x_l}||_2 \,|\, \boldsymbol{x_k} \in B_k, \boldsymbol{x_l} \in B_l\} \\ and \quad dist(B_k, B_l) &> (r_k + r_l), k \neq l; \quad k, l \in [1, K] \end{aligned} \qquad (6)$$

$$V_k(t) \leq V_{k,Max} \qquad (7)$$

In the above optimization problem, the constraints are defined for preventing collisions among UAVs. Mainly the constraints aim to fix the values of the deviation angles $\alpha_k(t)$ compared to the original velocity slope angles $\theta_k(t)$. On the other hand, the objective function (1) aims to minimize the deviation angles ($\alpha_k(t)$) of the UAVs' velocity vectors on their original paths ($\theta_k(t)$). This is in order to reduce the extra traveled distance to the original path. This will help in saving energy consumption and reducing the delay of the missions. In the optimization, constraints (3) and (4) provide the positions of UAVs at the next interval of time $(t+1)$ for a given set of actions. Constraint (5) defines the Euclidean ball $B_k$ associated with the error circle of a UAV k for the next candidate location. The distance between each point inside the ball and its limit should be inferior or equal to the UAV's radius. In what concern the constraint (6), it ensures that the Euclidean distance between every two points from two different Euclidean balls should be larger than the sum of their radii. We refer to this distance as the *safety distance*. This constraint prevents the collisions among UAVs by keeping the distance between them bigger than the safety distance. Finally, constraints (2) and (7) ensure that the proposed solution is realistic by limiting the velocities and deviation angles of UAVs. In fact, each UAV has a limited velocity according to different parameters. In this model, we consider that UAVs can deviate within a $2\pi$ interval, though the angle could be limited according to the nature of the UAVs used.

TABLE I
MODEL PARAMETERS.

| Notation | Description |
|---|---|
| $X$ | Geographical area width. |
| $Y$ | Geographical area height. |
| $X_k(t)$ | The candidate x-axis variable at the time $t$ of the UAV $k$. |
| $Y_k(t)$ | The candidate y-axis variable at the time $t$ of the UAV $k$. |
| $V_k(t)$ | The module for the expected velocity of the UAV $k$ at the time $t$. |
| $\theta_k(t)$ | The angle between the x-axis and the velocity vector of the UAV $k$, $V_k(t)$ at the time $t$. |
| $\alpha_k(t)$ | The deviation angle of the UAV $k$ comparing to $\theta_k(t)$ at the time $t$. |
| $r_k$ | The GPS error value of the UAV $k$. |
| $V_{k,Max}$ | The maximum velocity of the UAV $k$. |
| $V_{k,X}(t)$ | The candidate x-axis velocity vector variable at the time $t$ of the UAV $k$. |
| $V_{k,Y}(t)$ | The candidate y-axis velocity variable at the time $t$ of the UAV $k$. |

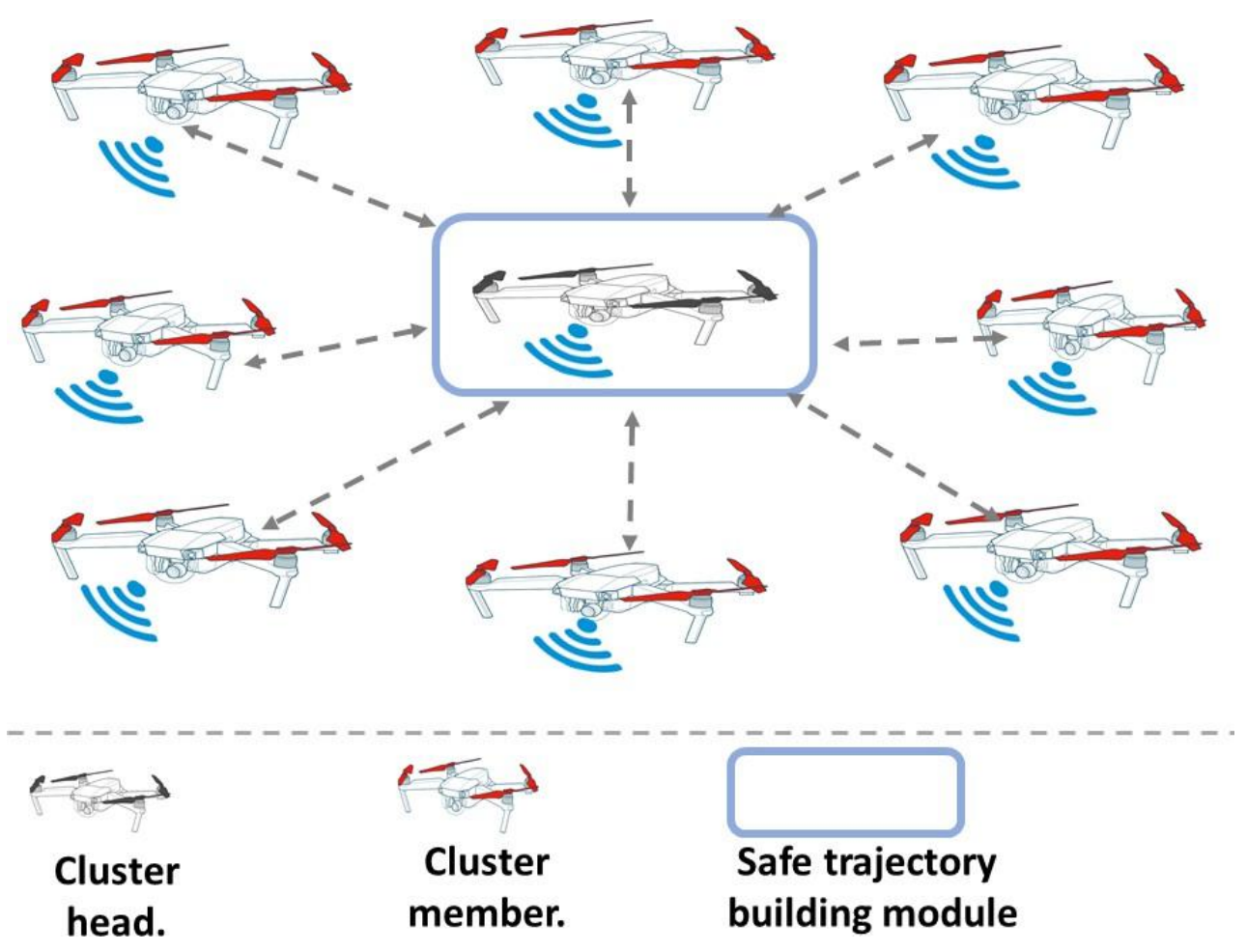

Fig. 2. UAVs' network architecture.

Since the Hessian matrices of constraints (3) and (4) are not positive semi-definite, we can conclude that the problem formulated above does not have a global minimum (i.e., Convex). Moreover, constraint (6) is not an affine convex function. Consequently, we need to reformulate the problem to find an optimal solution. Otherwise only a heuristic method can be used to solve it.

## III. EDC-UAV: Energy and Delay aware collision avoidance solution

This section focuses on the description of the multi-UAVs system architecture that is envisioned to implement the proposed solution. Also, this section aims to describe how the proposed solution avoids collisions among different UAVs.

### A. Solution overview

As depicted in Fig. 2, the proposed solution adopts a cluster-based topology. We consider that the same mission is shared among the swarm of UAVs, where one of them is elected as the cluster head (CH). In EDC-UAV, the CH is selected based on one of the following criteria as defined in [28]: *i*) the UAV that has the highest amount of energy supply; or *ii*) the UAV that has the shortest response time (i.e., highest computational capacity). The choice of the CH election strategy is driven by the nature of the mission. In fact, the first strategy is suitable for missions requiring a huge amount of computation, whereas the second strategy is used in missions that need quick response time. Taking into account the capacity limitations of UAVs in terms of energy and computation, we aim to endow UAVs with some autonomy that strengthens their safety by adopting a cluster-based topology. Moreover, the implementation of the proposed solution at the CH level helps in making fast decisions in preventing possible collisions.

The proposed solution is implemented as a secure trajectory building module that should be loaded on the CH. This module divides the cluster's mission time into a set of equal time slots of $\Delta t$ duration, where $t$ indicates the time slot in the algorithm execution. As indicated previously, at a time granularity $\tau$ within $\Delta t$, the CH executes the proposed module at each time slot $t$ for finding the optimal locations in the next slot $t+1$. Those locations should reduce the energy consumption and end-to-end delay of the mission while preventing physical collisions among UAVs.

As depicted in Fig.3, during the mission's time, at each $\tau$ within each $\Delta t$, the safe trajectory building module starts by analyzing its saved mission's details (e.g., mission's start position, mission's final position, and current GPS) received from the UAVs. If the current time $t$ equals the closing time $t_F$ , the module execution is stopped, which means that the mission is achieved. Otherwise, the safe trajectory building module calls the optimizer. The latter should compute for each UAV, the next location, the required velocity as well as the appropriate deviation angle which guarantees a safe move of UAVs without risk of collision. While the CH sends the different optimal next locations to each participating UAV

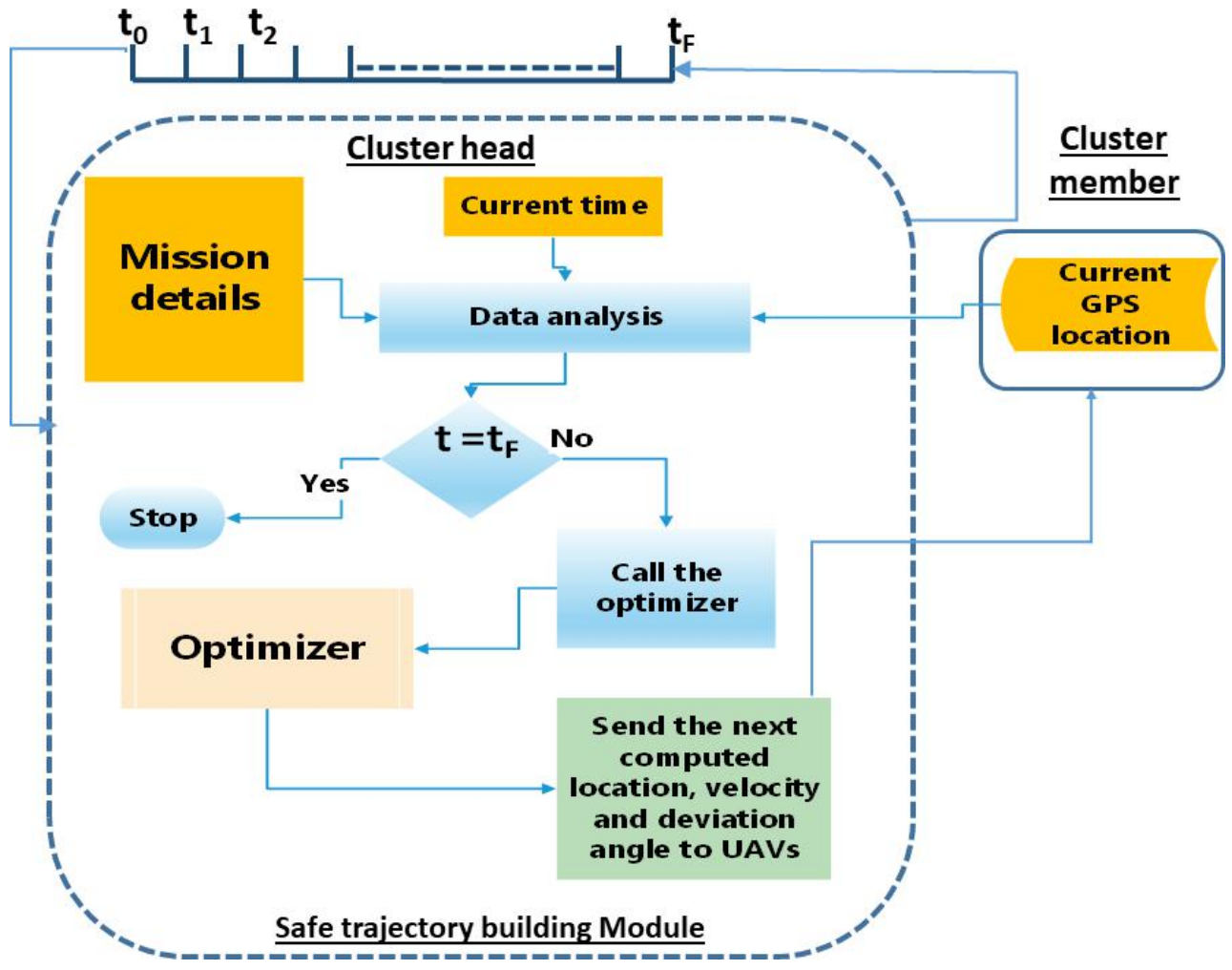

Fig. 3. Description of the safe trajectory building module.

in the cluster, the UAVs in turn will update their positions accordingly.

*B. Detailed solution description*

As explained above in Section II, the optimization problem, defined by Equations (1)-(7), is not convex. In this subsection, we reformulate the optimization problem so it becomes convex, and then we can solve it by finding the global optimum. The proposed optimization, herein, will be executed at the CH for finding the optimal paths. For this reason, we update the previous model by proposing a new objective function and modifying some constrains. The new model defines the location coordinates of each UAV $k$ at a given instant $t+1$, in terms of its location at the previous time $t$ as well as the velocity of each UAV at both axes: x-axis (i.e., $V_{k,X}(t)$) and y-axis ($V_{k,Y}(t)$). Formally, the optimization problem will set for each UAV $k \in \{1, 2, \ldots K\}$, the following items at each time $t \in \{1, 2, \ldots F\}$ : *i*) the velocity at x-axis ($V_{k,X}(t)$) and y-axis ($V_{k,Y}(t)$); *ii*) the location of UAV $k$ by defining its x-axis ($X_k(t)$) and its y-axis ($Y_k(t)$) coordinates. Formally, we have the following:

$$X_k(t+1) = X_k(t) + V_{k,X}(t) \times \Delta t \quad (8)$$

$$Y_k(t+1) = Y_k(t) + V_{k,Y}(t) \times \Delta t \quad (9)$$

The optimization problem that will be defined later should generate the values of $V_{k,X}(t)$, $V_{k,Y}(t)$, $X_k(t)$ and $Y_k(t)$. Based on those values, EDC-UAV can generate the velocity of UAV $k$ ($V_k(t)$) and its deviation angle $\alpha_k(t)$ at each time slot $t \in \{1, 2, \cdots F\}$ as follows:

$$V_k(t)^2 = V_{k,X}(t)^2 + V_{k,Y}(t)^2 \quad (10)$$

$$\alpha_k(t) = \tan^{-1}\left(\frac{E_Y(k) - Y_k(t)}{E_X(k) - X_k(t)}\right) \quad (11)$$

Moreover, $V_k(t)$ can be defined as follows:

$$V_k(t) = \frac{V_{k,X}(t)}{\cos(\alpha_k(t))} \quad (12)$$

In what follows, we define the optimization problem of EDC-UAV as follows:

$$\min \sum_{k=1}^{K}\sum_{t=1}^{F} (X_k(t) - X_k(t-1))^2 + (Y_k(t) - Y_k(t-1))^2 \quad (13)$$

$$s.t.$$

$\forall k \in [1, \cdots K], \forall t \in [1, \cdots F]$ :

$$X_k(t) = X_k(t-1) + V_{k,X}(t) \times \Delta t \quad (14)$$

$$Y_k(t) = Y_k(t-1) + V_{k,Y}(t) \times \Delta t \quad (15)$$

$\forall k, l \in [1, \cdots K], k \neq l, \forall t \in [1, \cdots F]$ :

$$(X_k(t) - X_l(t))^2 + (Y_k(t) - Y_l(t))^2 > (r_k + r_l)^2 \quad (16)$$

$\forall k \in [1, \cdots, K]$ :

$$Y_k(0) = S_Y(k) \quad (17)$$

$$X_k(0) = S_X(k) \quad (18)$$

$$X_k(F) = E_X(k) \quad (19)$$

$$Y_k(F) = E_Y(k) \quad (20)$$

$$(V_{k,X}(t))^2 + (V_{k,Y}(t))^2 = (\|V_k(t)\|)^2 \leq V_{k,Max}^2 \quad (21)$$

where $(S_X(k), S_Y(k))$ and $(E_X(k), E_Y(k))$ denote, respectively, the start and the end points of UAV $k$.

In the optimization problem of EDC-UAV, we aim to minimize, as much as possible, the new path distance of each UAV in order to prevent any physical collision. This new path distance minimization will also help in saving energy consumption and reducing response time of the UAVs. Formally, this can be achieved by reducing the inter-location distance at each instant $t$. Meanwhile, the constraints will ensure the following. Constraints 14 and 15 compute the x-axis and the y-axis coordinates at instant $t$ using the previous location and velocity, respectively. Constraint (16) prevents collisions between UAVs. Meanwhile, Constraints (17) – (20) ensure that each UAV should start from the initial point and stop at the end point. Last but not least, constraint (21) ensures that a UAV's velocity should not exceed its maximum allowed velocity. Mainly, this model aims to seek for the optimal velocity and direction by finding out $X^k(t), Y^k(t)$ that help in reducing the traveled distance while preventing collisions among UAVs.

The above optimization problem is still not convex due to constraint (16). In order to simplify the solution and make it convex, the following transformations are applied to constraint (16). Let $X'^{k,l}(t)$ and $Y'^{k,l}(t)$ denote two real variables, for $k, l \in [1, \cdots, K]$. Thus, the following constraints are added: $\forall t \in [1, \cdots, F]$, $\forall k, l \in [1, \cdots, K], k \neq l$ :

$$X'_{k,l}(t) \leq X_k(t) - X_l(t) \quad (22)$$

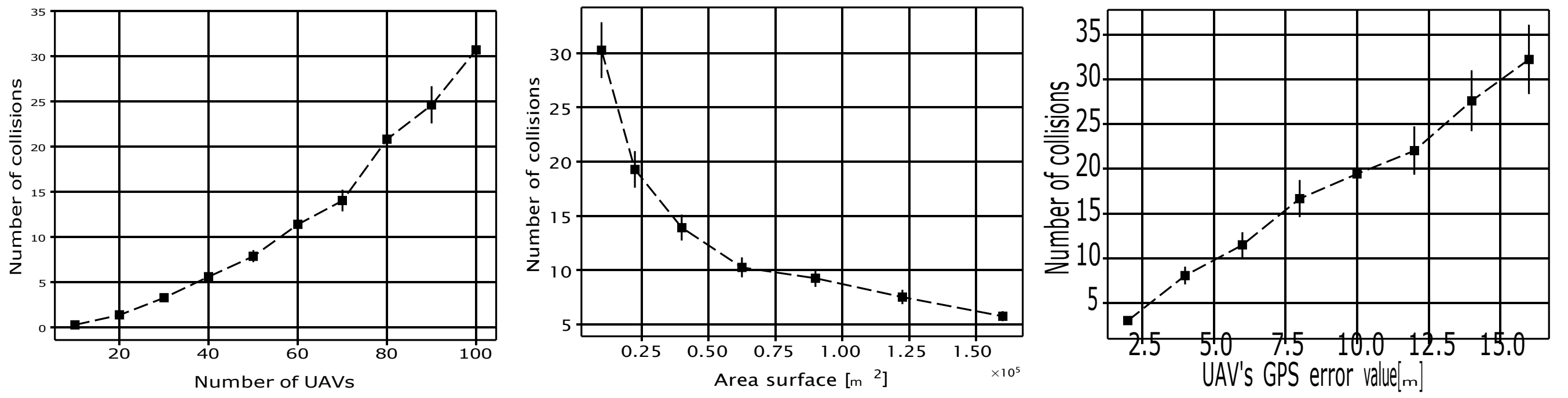

(a) Number of collisions vs Number of UAVs (b) Number of collisions vs Area surface (c) Number of collisions vs UAV's GPS error value

Fig. 4. Solution performance in terms of collision avoidance.

$$\dot{X}_{k,l}(t) \le X_l(t) - X_k(t) \quad (23)$$

$$\dot{Y}_{k,l}(t) \le Y_k(t) - Y_l(t) \quad (24)$$

$$\dot{Y}_{k,l}(t) \le Y_l(t) - Y_k(t) \quad (25)$$

$$-\dot{X}_{k,l}(t) - \dot{Y}_{k,l}(t) \ge (r_k + r_l)^2 \quad (26)$$

$$-\dot{X}_{k,l}(t) \ge 1 \quad (27)$$

$$-\dot{Y}_{k,l}(t) \ge 1 \quad (28)$$

The above constraints ensure the following. Constraints (22) and (23) ensure that: $-\dot{X}_{k,l}(t) \ge |X_k(t) - X_l(t)|$ while constraints (24) and (25) ensure that $-\dot{Y}_{k,l}(t) \ge |Y_k(t) - Y_l(t)|$. Constraint (26) ensures that $|X_k(t) - X_l(t)| + |Y_k(t) - Y_l(t)| \ge (r_k + r_l)^2$, while constraints (26), (27) and (28) help for ensuring that:

$$(X_k(t) - X_l(t))^2 + (Y_k(t) - Y_l(t))^2 \ge |X_k(t) - X_l(t)| + |Y_k(t) - Y_l(t)|$$
$$;and \quad |X_k(t) - X_l(t)| + |Y_k(t) - Y_l(t)| \ge (r_k + r_l)^2.$$

As a result, the previous model can be updated as follows:

$$min \sum_{k=1}^{K} \sum_{t=1}^{F} (X_k(t) - X_k(t-1))^2 + (Y_k(t) - Y_k(t-1))^2 \quad (29)$$

$s.t.$

$\forall k \in [1, \cdots K], \forall t \in [1, \cdots F]$ :

$$X_k(t) = X_k(t-1) + V_{k,X}(t) \times \Delta t \quad (30)$$

$$Y_k(t) = Y_k(t-1) + V_{k,Y}(t) \times \Delta t \quad (31)$$

$\forall k \in [1, \cdots K]$ :

$$Y_k(0) = S_Y(k) \quad (32)$$

$$X_k(0) = S_X(k) \quad (33)$$

$$X_k(F) = E_X(k) \quad (34)$$

$$Y_k(F) = E_Y(k) \quad (35)$$

$$(\|V_{k,X}(t)\|)^2 + (\|V_{k,Y}(t)\|)^2 = (\|V_k(t)\|)^2 \le V_{k,Max}^2 \quad (36)$$

$\forall t \in [1, \cdots F] \forall k, l \in [1, \cdots K], k \neq l$ :

$$\dot{X}_{k,l}(t) \le X_k(t) - X_l(t) \quad (37)$$

$$\dot{X}_{k,l}(t) \le X_l(t) - X_k(t) \quad (38)$$

$$\dot{Y}_{k,l}(t) \le Y_k(t) - Y_l(t) \quad (39)$$

$$\dot{Y}_{k,l}(t) \le Y_l(t) - Y_k(t) \quad (40)$$

$$-\dot{X}_{k,l}(t) - \dot{Y}_{k,l}(t) \ge (r_k + r_l)^2 \quad (41)$$

$$-\dot{X}_{k,l}(t) \ge 1 \quad (42)$$

$$-\dot{Y}_{k,l}(t) \ge 1 \quad (43)$$

## IV. Simulation

Throughout this section, we will assess the performance of the EDC-UAV protocol. The Gurobi optimizer [29] is used to solve the proposed optimization model, while the performance evaluation is performed through simulation. The EDC-UAV protocol is evaluated based on the following metrics: *i*) the number of collisions; *ii*) the extra distance traveled; and *iii*) the execution time.

In the simulation scenarios, a two-dimensional horizontal (2D-H) area is considered. Each UAV is modeled as a point within a circle which represents the margin error of the GPS sensor. It is assumed that the actual position of the UAV might be any point that lies inside this circle.

To evaluate the performance mentioned above, the following parameters are considered in our experiments: *i*) the number of UAVs; *ii*) the area surface; and *iii*) the UAV's GPS error. When the number of UAVs is swept, the area surface is set to 10000 $m^2$ and the UAV's GPS error is set to 5 $m$. When the area surface is swept, the number of UAVs is set to 50 UAVs and the UAV's GPS error is set to 5 $m$. When the UAV's GPS error is swept, the number of UAVs is set to 50 and the area surface is set to 40000 $m^2$.

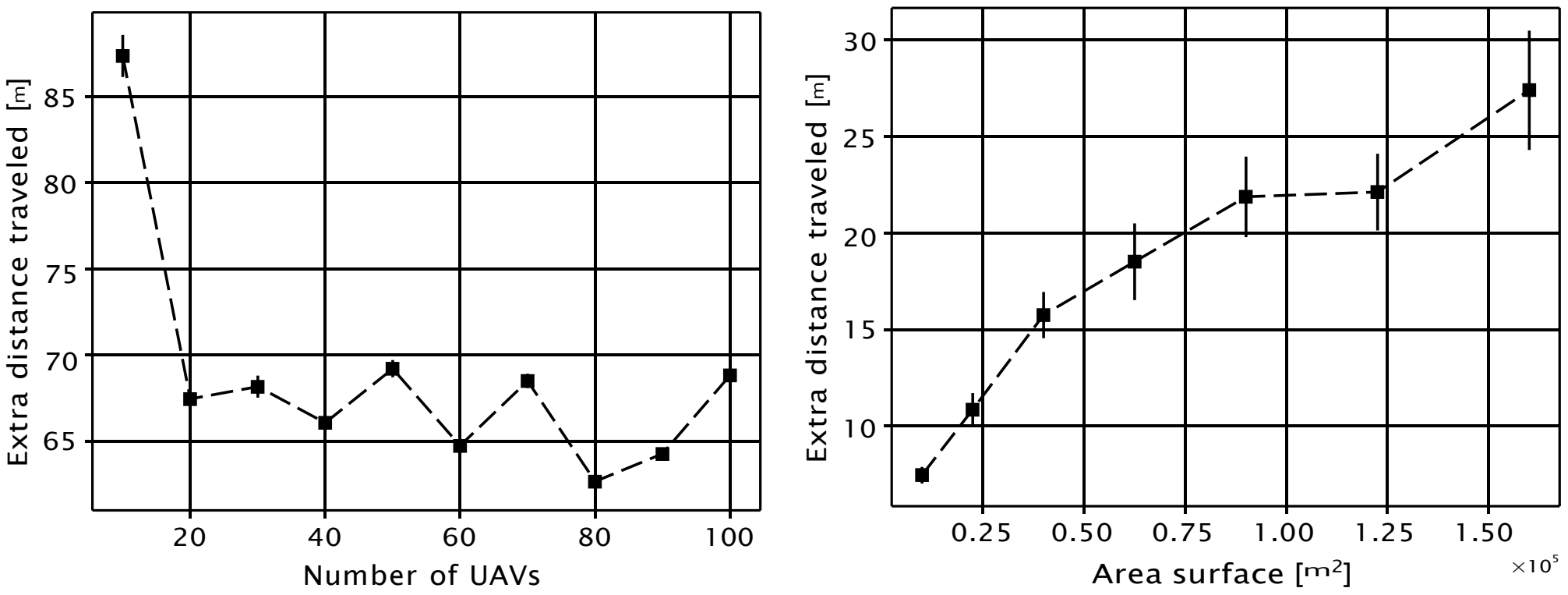

(a) Extra distance traveled vs number of UAVs.

(b) Extra distance traveled vs area surface.

Fig. 5. Solution performances in terms of extra traveled distance.

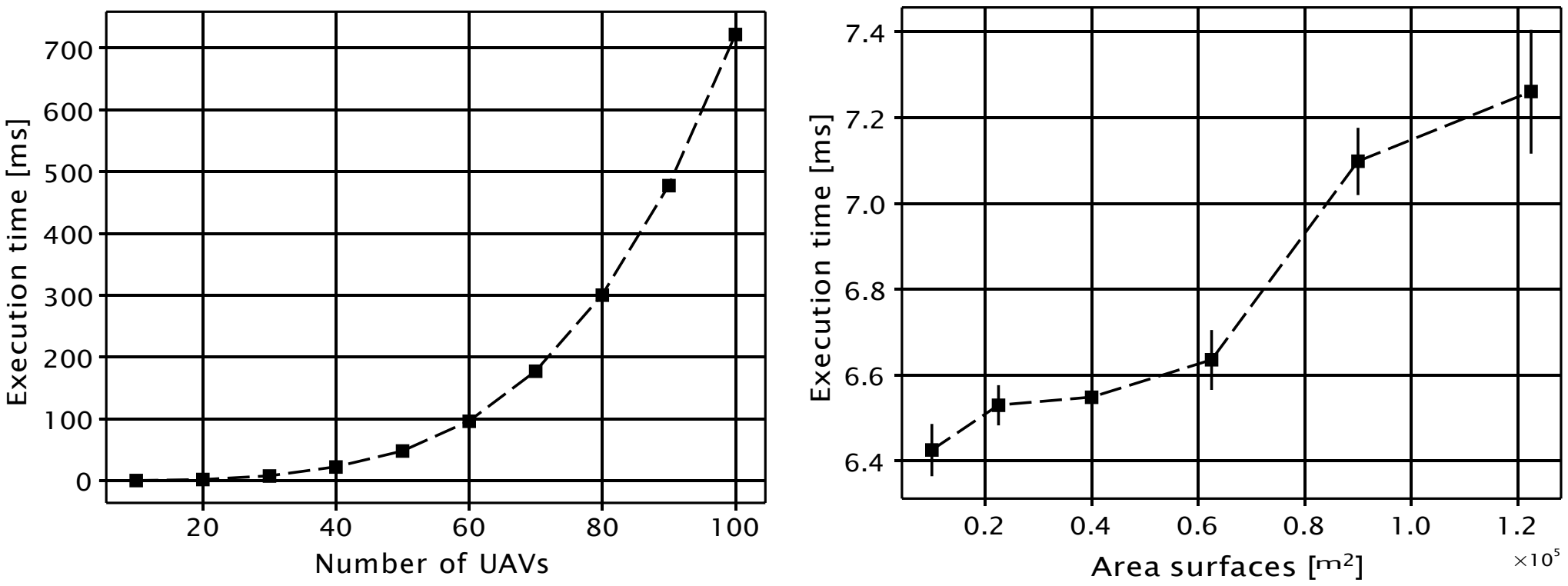

(a) Execution time vs number of UAVs.

(b) Execution time vs area surface

Fig. 6. Solution performances in term of execution time.

### *A. Number of collisions*

Figures 4(a), 4(b), and 4(b) show, respectively, the number of collisions versus the number of UAVs, the area surface, and the UAV's GPS error when the EDC-UAV collision avoidance mechanism is disabled. In these figures, each point was generated by averaging 30 independent runs. A potential collision is detected whenever the bounding circles, representing the possible locations of two or more UAVs, overlap.

As expected, the results indicate that both the number of UAVs and the UAV's GPS error value have a negative impact on the number of collisions, i.e., the higher their values are, the more significant number of collisions among UAVs will be. Conversely, an increase in the surface area value reduces the number of possible collisions. Our results show that the number of collisions depends quadratically on the UAVs density, whereas it exhibits a linear dependence on the UAV's GPS error.

These results motivate the use of collision avoidance solutions like EDC-UAV to guarantee that UAVs accomplish their missions in safe conditions, especially in scenarios with a high density of UAVs.

### *B. Extra distance traveled*

Figures 5(a) and 5(b) show respectively the average extra distance traveled by all UAVs as a function of the number of UAVs and the area surface. Note that the longer the extra distance traveled, the higher the energy consumption. Unexpectedly, the simulation results (see Fig. 5(a)) shows that the number of UAVs has a positive impact on the extra distance traveled. The more UAVs in the cluster, the less extra distance covered.These results could be explained by the fact that the proposed linear programming optimization model aims to minimize as much as possible the distance between each consecutive UAV's location (see constraint (29)). However, the simulation results show that the area surface has a negative impact on the extra distance traveled (see Fig. 5(b)). We notice that in the simulated scenario the UAVs' trajectories are constructed aleatory following a normal distribution. The wider the surface area, the bigger the probability of having a wide distance between the UAV's start and final positions.

### *C. Execution time*

Figures 6(a) and 6(b) depict respectively the execution time of EDC-UAV as a function of the number of UAVs and the surface area. The simulation results show that an increase in the number of UAVs impacts the EDC-UAV execution time negatively (see Fig. 6(a)). As the number of UAVs in the network increases, EDC-UAV spends more time computing the UAVs' trajectories. Furthermore, the simulation results show that the expansion of the surface area also escalates EDC-UAV execution time (see Fig.6(b)). As noticed previously, the higher the surface area, the higher the probability of having a wide distance between the start and the final locations of the UAVs.

## V. Conclusion

Collision avoidance remains a big challenge for real integration of UAVs in different applications in spite of the numerous solutions that have been proposed towards addressing it. In this paper, we present the EDC-UAV solution that aims to avoid physical collisions among a cluster of UAVs by building consecutively, in flight, their optimal safe trajectories. The optimization in EDC-UAV resolves, periodically, a linear programming model that seeks to find the next optimal UAVs' localizations that ensure their safety, saves their energy consumption as well as minimizes their execution time. The simulation results show the efficiency of the proposed protocol. The obtained results motivate the necessity of the proposed protocol to enhance the safety of the UAVs while saving their energy consumption.